\documentclass[aip,apl,reprint]{revtex4-1}

\usepackage{graphicx}% Include figure files
\usepackage{dcolumn}% Align table columns on decimal point
\usepackage{bm, textcomp,amsmath,amssymb,epsfig}% bold math
\begin{document}

\title[APPLIED PHYSICS LETTERS xx, xxxx (2011)]{Electrical spin injection and detection in Germanium using three terminal geometry}% Force line breaks with \\
%\thanks{Footnote to title of article.}

\author{A. Jain}
\author{L. Louahadj}
\affiliation {SP2M, CEA/UJF Grenoble 1, INAC, F-38054, Grenoble, France}
\author{J. Peiro}
\author{J. C. Le Breton}
\affiliation {Unit\'e Mixte de Physique CNRS-Thal\`es, 91767 Palaiseau et Universit\'e Paris-Sud, 91405, Orsay, France}
\author{C. Vergnaud}
\author{A. Barski}
\author{C. Beign\'e}
\author{L. Notin}
\author{A. Marty}
\affiliation {SP2M, CEA/UJF Grenoble 1, INAC, F-38054, Grenoble, France}

\author{V. Baltz}
\author{S. Auffret}
\affiliation {SPINTEC, UMR CEA/CNRS/UJF Grenoble 1/Grenoble-INP, INAC, F-38054, Grenoble, France}

\author{E. Augendre}
\affiliation {CEA, LETI, MINATEC Campus, F-38054, Grenoble, France}

\author{H. Jaffr\`es}
\author{J. M. George}
\affiliation {Unit\'e Mixte de Physique CNRS-Thal\`es, 91767 Palaiseau et Universit\'e Paris-Sud, 91405, Orsay, France}
\author{M. Jamet}
 \email{matthieu.jamet@cea.fr}
 \affiliation {SP2M, CEA/UJF Grenoble 1, INAC, F-38054, Grenoble, France}
 
\date{\today} 

\begin{abstract} 

In this letter, we report on successful electrical spin injection and detection in \textit{n}-type germanium-on-insulator (GOI) using a Co/Py/Al$_{2}$O$_{3}$ spin injector and 3-terminal non-local measurements. We observe an enhanced spin accumulation signal of the order of 1 meV consistent with the sequential tunneling process via interface states in the vicinity of the Al$_{2}$O$_{3}$/Ge interface. This spin signal is further observable up to 220 K. Moreover, the presence of a strong \textit{inverted} Hanle effect points at the influence of random fields arising from interface roughness on the injected spins.

\end{abstract}

%\pacs{75.50.Pp, 75.75.-c, 75.30.Gw, 61.46.-w, 76.30.-v}

\maketitle

Semiconductor devices like spin-FETs based on spin currents are highly desirable because of their high performance, increased functionality and low power consumption\cite{Wolf2001,Sugahara2004}. Recently germanium is gaining huge interest in semiconductor-spintronics industry due to its large spin-diffusion length attributed to the inversion symmetric crystal structure and high carrier mobility\cite{Dyakonov2008}. In order to obtain spin-based devices, electrical spin injection and detection are the key factors to be considered. The successful demonstration of electrical spin injection and detection have been shown in Si \cite{Jonker2007,Appelbaum2007,Grenet2009,Dash2009} and GaAs \cite{Lou2007} but little advance has been done in the case of germanium yet. Liu \textit{et al.} demonstrated electrical spin injection in Ge nanowires using Co/MgO contacts and reported a spin diffusion length of more than 100 $\mu$m at 4.5 K \cite{Liu2010}. Zhou \textit{et al.} reported electrical spin injection and detection in bulk Ge using epitaxially grown Fe/MgO on $n$-Ge in 4-contact non-local geometry and found spin lifetimes as long as 1 ns at 4 K\cite{Zhou2011}. Recently Saito \textit{et al.} reported on electrical spin injection and detection in \textit{p}-type germanium also using Fe/MgO\cite{Saito2011}. In this letter we demonstrate the electrical spin injection and detection in \textit{n}-type germanium using Al$_{2}$O$_{3}$ tunnel barrier in 3-terminal geometry\cite{Dash2009,Li2011,Tran2009}. The enhanced spin accumulation signal ($\Delta V$=0.5 mV) as compared to theoretical predictions is strong indication that spin accumulation rather occurs on localized states at the Al$_{2}$O$_{3}$/Ge interface. Finally we study the effect of interface roughness on the spin polarization. The experiments presented here were carried out on doped germanium-on-insulator substrates (GOI). These substrates were fabricated using the Smart Cut$^{TM}$ process and Ge epitaxial wafers\cite{Deguet2005}. The transferred 40 nm-thick Ge film was $n$-type doped in two steps: a first step (phosphorus, 3x10$^{13}$ cm$^{-2}$, 40 keV, annealed for 1h at $550^{\circ}$C) that provided uniform doping in the range of 10$^{18}$ cm$^{-3}$ (resistivity $\rho$=10 m$\Omega$.cm), and a second step (phosphorus, 2x10$^{14}$ cm$^{-2}$, 3 keV, annealed for 10 s at $550^{\circ}$C) that increased surface $n^{+}$ doping to the vicinity of 10$^{19}$cm$^{-3}$. The thickness of the \textit{n$^{+}$}-doped layer is estimated to be 10 nm. The surface of the GOI was finally capped with amorphous SiO$_{2}$ to prevent surface oxidation.

\begin{figure}[hbt!!]
\begin{center}
\begin{tabular}{cc}
\mbox{\includegraphics[angle=0,width=0.25\textwidth]{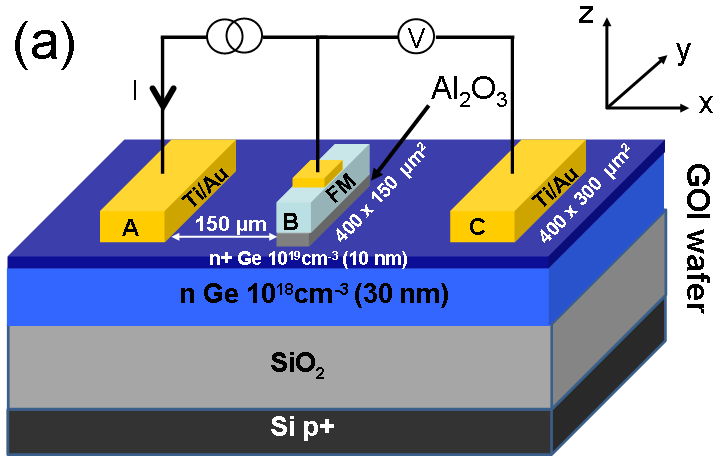}}&
\mbox{\includegraphics[angle=0,width=0.23\textwidth]{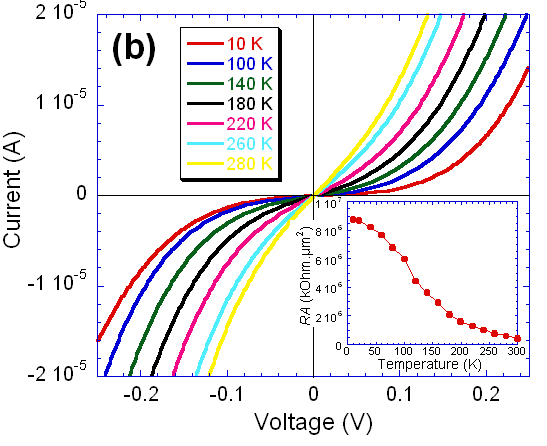}}
\end{tabular}
\caption{(a) Schematic drawing of the three-terminal device used for electrical spin injection and detection in germanium. Magnetic field is applied either along $y$ (in-plane geometry) or along $z$ (out-of-plane geometry). (b) $I-V$ characteristics of the Pt/Co/NiFe/Al$_{2}$O$_{3}$ tunnel contact at various temperatures. Inset: temperature dependence of the tunnel contact $RA$ product ($I$=0.5 $\mu$A). }
\end{center}
\end{figure}

In Ge, there is a strong Fermi-level pinning close to the valence band edge which leads to high Schottky barrier height (SBH=0.6 eV) of the order of the bandgap itself and large depletion width for metal/\textit{n}-Ge interface\cite{Nishimira2008}. By inserting a thin Al$_{2}$O$_{3}$ layer, we drastically reduce the SBH below 0.3 eV\cite{Zhou2008}. In our specific case, the $n^{+}$ surface doping layer sharply reduces the thickness of the Schottky barrier with the result that the tunneling transparency is enhanced. The GOI substrates were treated with hydrofluoric solution to remove the SiO$_{2}$ capping layer and introduced in the sputtering machine. Aluminium layer of 1.6 nm was grown and treated with oxygen plasma to form alumina barrier. Then stack of 5 nm permalloy, 20 nm Co and 10 nm Pt was grown in order to have in-plane anisotropy with coercive field of 10 Oe. The insertion of alumina layer alleviates the Fermi level pinning and acts as tunnel barrier for efficient spin injection\cite{Fert2001}. The sample was processed using standard optical lithography and dry etching to have 150$\times$400 $\mu$m$^{2}$ magnetic electrodes. Finally ohmic contacts of Ti/Au with dimensions 300$\times$400 $\mu$m$^{2}$ were deposited to form three-terminal geometry. The schematic diagram of the structure is shown in Fig. 1a. Fig. 1b displays the $I-V$ characteristics between the tunnel contact B and one ohmic contact. The behavior is highly non-linear showing up a tunneling-like transmission and only slightly dependent on temperature. In the inset of Fig. 1b, we indeed find: $R(10K)/R(300K)\approx$8 for a DC current of 0.5 $\mu$A. This proves that the tunneling process through the thick Al$_{2}$O$_{3}$ barrier is the dominant mechanism for the transport. This is made possible by the specific shape of the Schottky barrier in Ge, thin enough to allow a tunneling transmission through it and low enough to limit its own resistance. Nevertheless one cannot rule out its role on a possible confinement effect for spins injected at the direct Al$_{2}$O$_{3}$/Ge interface. In the whole temperature range, the $RA$ product exceeds the minimum interface resistance threshold required for spin injection into Ge\cite{Fert2007}: ($\rho l_{sf}^{2})/w\approx$10 k$\Omega$.$\mu$m$^{2}$ where $w$=30 nm is the thickness of the channel and where $l_{sf}$ was taken to be of the order of 1 $\mu$m \cite{Zhou2011}. For non-local 3-terminal measurements, a constant current $I$ was passed between contacts A and B and a voltage V$_{BC}$ was measured between contacts B and C as a function of the external field \textbf{B}$^{ext}$. \textbf{B}$^{ext}$ is either out-of-plane along $z$ (Hanle effect) or in-plane along $y$ ($inverted$ Hanle effect). For $I>0$ (resp. $I<0$), electrons are injected in (resp. extracted from) the Ge film.

\begin{figure}[hbt!!]
\begin{center}
\begin{tabular}{cc}
\mbox{\includegraphics[angle=0,width=0.23\textwidth]{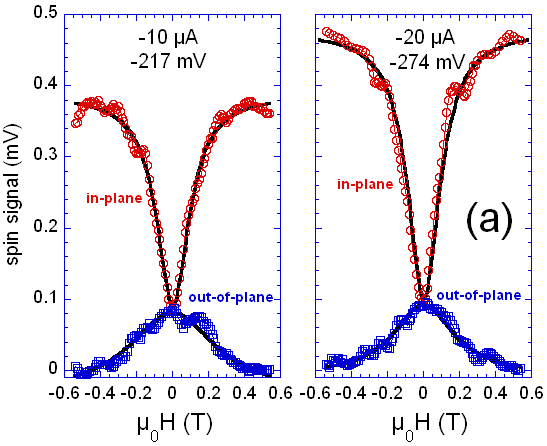}}&
\mbox{\includegraphics[angle=0,width=0.23\textwidth]{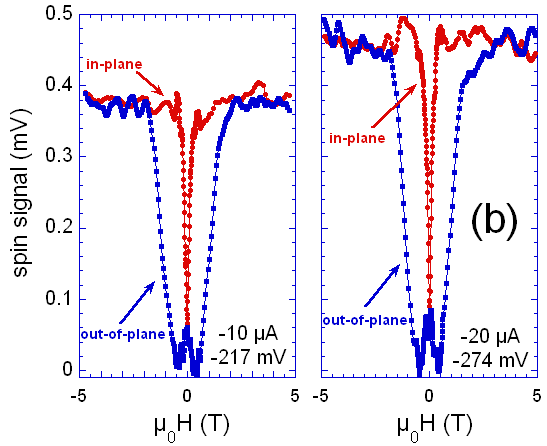}}\\
\mbox{\includegraphics[angle=0,width=0.23\textwidth]{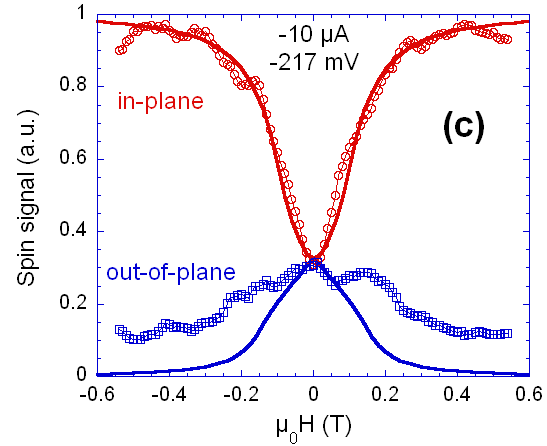}}&
\mbox{\includegraphics[angle=0,width=0.23\textwidth]{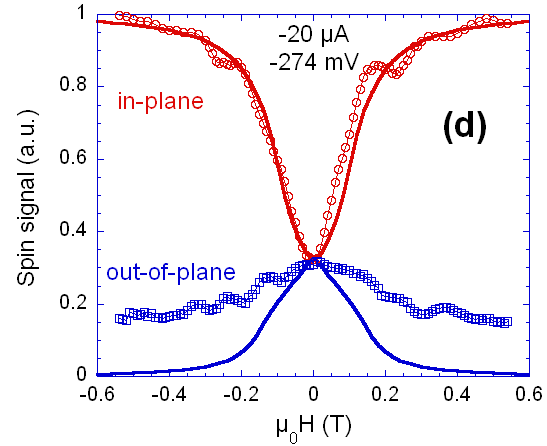}}
\end{tabular}
\caption{(a) low field and (b) high field dependence of the spin signal for two different bias currents -10 $\mu A$ (-217 mV) and -20 $\mu A$ (-274 mV) showing both Hanle (out-of-plane) and inverted Hanle (in-plane) effects. Measurements were performed at 10 K. Black solid lines in (a) are Lorentzian fits. (c), (d) Comparison between experimental (open symbols) and calculated (solid lines) spin signals for in-plane and out-of-plane configurations at -10 $\mu$A and -20 $\mu$A. Data were normalized to the maximum value.}
\end{center}
\end{figure}

Fig. 2a and 2b display Hanle curves in the low and high field regimes respectively. Measurements were carried out at 10 K for two different DC currents: -10 $\mu$A ($V$=-0.217 V) and -20 $\mu$A ($V$=-0.274 V). We observe a voltage drop V$_{BC}$ of 0.1 mV, hence providing evidence of spin accumulation and then spin injection in Ge. As a first approximation the Hanle curves can be described, at least for localized electrons, by a Lorentzian shape given by $\Delta V=\Delta V_{o}/(1+(\omega_{L}\tau_{sf})^{2})$ where $\tau_{sf}$ is the spin lifetime and $\omega_{L}$ is the Larmor frequency ($\omega_{L}=g\mu_{B}B_{z}/\hbar$, where $g$ is the Land\'e
factor ($g$=1.6 for Ge\cite{Feher1959}) and $\mu_{B}$ is the Bohr magneton. After fitting the Hanle curve, we get a spin lifetime of 35 ps which is much shorter than 1 ns reported by Zhou et al. in \textit{n}-type Ge at 4 K\cite{Zhou2011}. This may be explained by the local random magnetostatic fields \textbf{B}$^{ms}$ arising from finite interface roughness that severely reduce the spin accumulation. This phenomenon was recently observed in Si and GaAs and verified by \textit{inverted} Hanle effect\cite{Dash2011}.

\begin{figure}[hbt!!]
\begin{center}
\begin{tabular}{cc}
\mbox{\includegraphics[angle=0,width=0.23\textwidth]{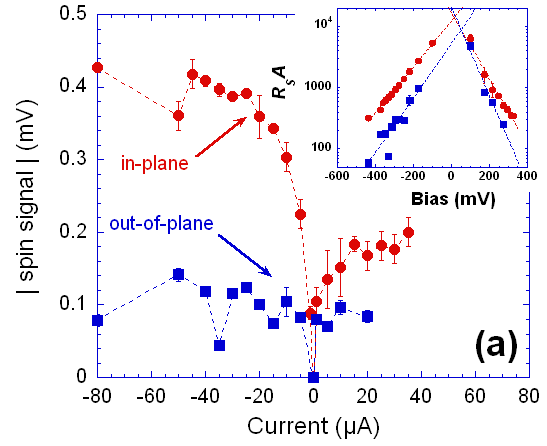}}&
\mbox{\includegraphics[angle=0,width=0.23\textwidth]{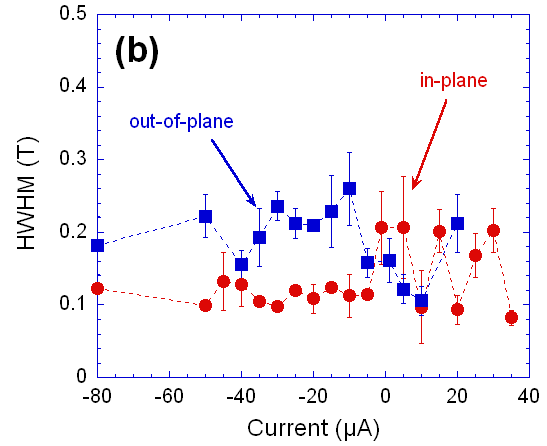}}\\
\mbox{\includegraphics[angle=0,width=0.23\textwidth]{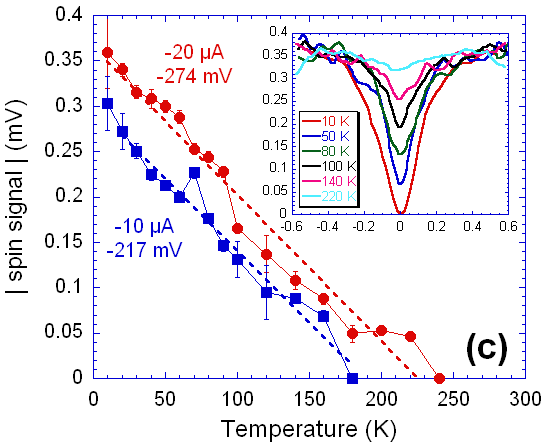}}&
\mbox{\includegraphics[angle=0,width=0.23\textwidth]{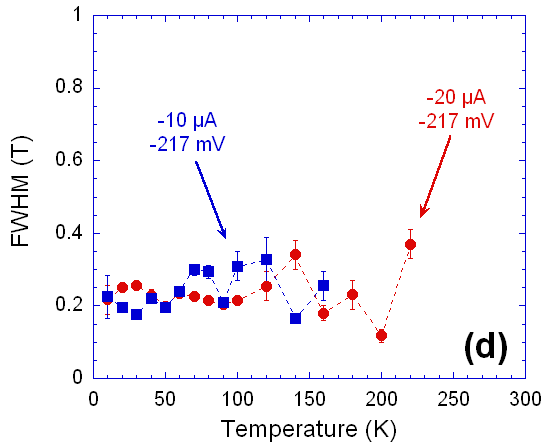}}
\end{tabular}
\caption{(a),(b) Bias dependence of the spin signal and of the half width at half maximum (HWHM) at 10 K for both in-plane and out-of-plane geometries. The inset in (a) shows the bias dependence of the spin resistance-area product $R_{s}.A$ in k$\Omega$.$\mu$m$^{2}$. Dashed lines are guides for the eye. (c) Temperature dependence of \textit{inverted} Hanle effect for two different bias currents -10 $\mu A$ (-217 mV) and -20 $\mu A$ (-274 mV). Spin signal is still observable at 220 K. The inset shows the inverted Hanle signal measured for a bias current of -20 $\mu A$ (-274 mV) at various temperatures. (d) Temperature dependence of the HWHM of Hanle curves for -10 $\mu A$ (-217 mV) and -20 $\mu A$ (-274 mV) bias currents.}
\end{center}
\end{figure}

For \textit{inverted} Hanle effect measurements, an external field \textbf{B}$^{ext}$ is applied in-plane along the magnetization direction to enhance the component of the effective field along the direction of the injected spins. The spin precession and decoherence are then gradually suppressed and the signal increases (in-plane curves in Fig. 2). At high fields, both Hanle and \textit{inverted} Hanle curves saturate and perfectly coincide above the cobalt demagnetizing field of 1.8 T. The total spin signal is thus the difference between the maximum in-plane and minimum out-of-plane values, we obtain: $\Delta V$=0.5 mV at 10 K and -20 $\mu$A. Note that the minimum out-of-plane value may not necessarily coincide with the total loss of spin accumulation since the magnetic moment of the electrode starts to align along the applied field when reaching a fraction of the demagnetizing field. Hence the total spin signal we measure represents a lower bound. In 3-terminal measurements, $\Delta V$=$\gamma$$\Delta\mu$/2$\left|e\right|$ where $e$ is the electron charge, $\gamma$=0.3 is the spin transmission coefficient through the alumina barrier\cite{Dash2009} and $\Delta\mu$=$\mu_{\uparrow}-\mu_{\downarrow}$ is the difference of electrochemical potentials for spin $up$ ($\uparrow$) and spin $down$ ($\downarrow$). The spin resistance-area product $R_{s}.A=(\Delta V/I).A=\gamma^{2}\rho l_{sf}^{2}/w\approx$1500 k$\Omega$.$\mu$m$^{2}$ where $A$ is the FM contact area is almost 4 orders of magnitude larger than the one expected by considering the spin diffusion length ($l_{sf}$=1 $\mu$m) reported by Zhou et al. in $n$-type Ge at 4 K\cite{Zhou2011}. This is a strong indication that spin injection through localized states (e.g. P donors in the depletion layer or surface states at the Al$_{2}$O$_{3}$/Ge interface) is at play\cite{Tran2009}. In order to estimate \textbf{B}$^{ms}$, we have performed atomic force microscopy (AFM) measurements on GOI wafers after alumina deposition. We indeed found a RMS roughness of 0.4 nm with a correlation length of the order of 45 nm. Then we have considered a regular array of magnetic charges with a period of 45 nm and calculated the three components ($B_{x}^{ms}$,$B_{y}^{ms}$,$B_{z}^{ms}$) of the magnetostatic field acting on injected spins. Spin dynamics has been computed by considering only spin precession and relaxation. Spin drift and diffusion were neglected as discussed in Ref. \cite{Dash2011}. The results are shown in Fig. 2c and 2d where the spin component along the FM magnetization is plotted as a function of in-plane and out-of-plane external fields. The following parameters were used: $\tau_{sf}$=1 ns \cite{Zhou2011} (note that any spin lifetime longer than 1 ns leads to the same results) and $\mu_{0}M_{s}$=0.9 T. Moreover we found the best agreement with experimental curves at a depth of 6 nm away from the FM/Al$_{2}$O$_{3}$ interface \textit{i.e.} 3-4 nm deep in the Ge layer. The agreement with \textit{inverted} Hanle effect is very good whereas for Hanle measurements the spin signal seems not to reach its minimum value probably because the FM magnetization starts to rotate out-of-plane at low field as discussed previously. We finally studied the evolution of spin signals as a function of DC current and temperature. As shown in Fig. 3a, in-plane and out-of-plane signals are almost symmetric with respect to zero bias except a factor 2 between spin injection and spin extraction regimes. In the inset of Fig. 3a, the spin resistance-area product $R_{s}.A$ is displayed (in logarihmic scale) as a function of the bias voltage. It clearly decreases exponentially for positive and negative bias voltage. However the slopes are different and this effect may be explained by the asymmetric modulation of the depletion width and then Schottky resistance with the bias. This bias dependence seems to indicate that a low density of interfacial states is likely to form a band extending in the whole Ge bandgap. The low 2-D density of states associated to such interfacial states should then give rise to a correlated high $R_{s}.A$ value as observed experimentally.
In Fig. 3c, the spin signal decreases almost linearly with temperature. This behavior has already been observed by Li \textit{et al.} in silicon \cite{Li2011} and the origin of this linear variation is still under investigation but could be related to the leakage of the Schottky resistance as a function of temperature. Most remarkable is that we still observe spin signal up to 220 K. On the other hand, in Fig. 3b and 3d we can notice that the HWHM of Hanle and \textit{inverted} Hanle curves is almost constant with DC current and temperature. This behavior clearly supports our assertion that the Hanle curve broadening leading to an underestimation of spin lifetime and \textit{inverted} Hanle effect are not due to intrinsic spin relaxation mechanisms but rather to random magnetostatic fields arising from interface roughness. Nevertheless one cannot totally rule out possible spin decoherence through hyperfine interaction with localized nuclear spins on Ge atoms.\\ 
To summarize, we have successfully created and detected spin accumulation in $n$-type Ge. The enhanced spin signal as compared to theoretical values seems to indicate that spin accumulation actually occurs on localized states close to the Al$_{2}$O$_{3}$/Ge interface. Moreover the underestimated spin lifetime and the observation of $inverted$ Hanle effect are consistent with spin dephasing due to random magnetostatic fields arising from interface roughness. This is further supported by the constant HWHM of Hanle and \textit{inverted} Hanle curves with DC current and temperature. Finally this spin accumulation signal could be detected up to 220 K.\\
This work was granted by the Nanoscience Foundation of Grenoble (RTRA project IMAGE). The initial GOI substrates were obtained through collaboration with Soitec under the public funded NanoSmart program (French OSEO).

\end{document}